\documentstyle[12pt]{article}
\begin{document}
\setlength{\baselineskip}{0.30in}

\newcommand{\beq}{\begin{equation}}
\newcommand{\eeq}{\end{equation}}

\newcommand{\bi}{\bibitem}
\def\mpl{m_{Pl}}
\begin{flushright}
UM-TH-94-20\\
DPNU-94-23\\
June, 1994\\
hep-ph/9406399
\end{flushright}

\begin{center}
\vglue .06in
{\Large \bf {Nambu-Jona-Lasinio Model vs QCD}}
\\[.5in]
{\bf Koichi Yamawaki}
\\[.05in]
{\it Department of Physics\\
Nagoya University\\
Nagoya, 464-01, Japan}\\
{\bf and}\\
{\bf V.I. Zakharov}\footnote{On leave of absence from Max-Planck Institute
for Physics, Munich}\\
{\it{The Randall Laboratory of Physics\\
University of Michigan\\
Ann Arbor, MI 48109}}\\[.15in]
\end{center}
\begin{abstract}
\begin{quotation}
We argue that the extended Nambu-Jona-Lasinio model
which introduces four-quark interactions to describe
hadron physics at low energies can be confronted with
fundamental quantum chromodynamics by means of spectral sum rules.
While there arise rather restrictive bounds
on the strength of the effective four-quark interaction in the vector
channel, introduction of the four-quark interaction in the pseudoscalar
channel resolves a long standing puzzle of the sum rules.

\end{quotation}
\end{abstract}
\newpage

More than thirty
 years ago Nambu and Jona-Lasinio (NJL) proposed a model of
superconductivity type as an effective theory of hadrons at low energies
\cite{nambu}.
The basic feature of the NJL model is the introduction of
effective four-fermion interaction.
In terms of quark interactions the
extended version of the NJL model
Lagrangian is
parameterized by two couplings $G_{S,V}$:
\beq
L_{\rm int}~=~{{G_S}\over 2}\left((\bar
q\lambda^{\alpha}q)^2+(\bar qi\gamma_5\lambda^{\alpha}q)^2\right)-
{{G_V}\over 2}\left((\bar q\gamma_{\mu}\lambda^{\alpha}q)^2+
(\bar q\gamma_{\mu}\gamma_5\lambda^{\alpha}q)^2\right),
\label{one}
\eeq
where $q$ are quark fields, $\lambda^{\alpha}$ are the Gell-Mann SU(3)
matrices in the flavor space and the color indices are suppressed.
Since (\ref{one}) is an effective Lagrangian, loop integrations
are allowed up to an ultraviolet cutoff $\Lambda_{UV}$ which
represents one more parameter. With time going on, the original idea
\cite{nambu} has developed into rich and successful phenomenology of
hadrons at low energy (for a review see, e.g., Ref.\cite{volkov}).
The modern way to confront the model with the data is to calculate
the parameters of pion interactions at low energies
\cite{bijnens}.

A crucial theoretical problem -- common to most of the models with
effective Lagrangians --
is whether interaction (\ref{one}) can be derived from QCD.
As usual, the effective Lagrangian should share the symmetries
of the fundamental one. If one confines oneself to four-quark
interaction, then the two terms in (\ref{one})
are the only ones consistent with the chiral invariance.
As for the four-fermionic form itself, it is mostly heuristic, although
attempts of its dynamical derivation have also been made,
see in particular review in Ref.\cite{ball}.
Generally speaking, the use of the
effective Lagrangian (\ref{one}) and of perturbative
calculations within the fundamental QCD is justified
in different kinematical regions. Namely, the language of (nearly)
massless
quarks and of gluons becomes rigorous at short distances where the
running
coupling is small, while the effective Lagrangian (\ref{one}) describes
the large-distance, or the low-momentum dynamics.

In this Letter we will attract attention to the possibility
of confronting NJL model and fundamental QCD directly,
where by fundamental QCD we will understand QCD sum rules \cite{SVZ}.
The QCD sum rules exploit perturbative calculations and push them to
as low momenta as possible. It was demonstrated that the sum rules
work down to Euclidean momenta $Q^2\sim 0.5 ~{\rm GeV}^2$. On the other
 hand, a fit
to the parameters of the NJL model looks as \cite{volkov}:
\beq
\Lambda_{UV}~\approx~1.25~{\rm GeV},~ G_S~\approx~5~ {\rm GeV}^{-2},
{}~G_V~\approx~10~{\rm GeV}^{-2}
\label{numbers}.
\eeq
The crucial observation is that $\Lambda_{UV}$ is in fact rather large.
As a result there exists so to say moderate $Q^2$,
\beq
0.5~{\rm GeV}^2~\le~ Q^2_{\rm moderate}~\le~ 1.5~{\rm GeV}^2
,\eeq
where the both approaches claim their validity. In this region one can
probe the NJL model via the QCD sum rules. The result
is that the value of $G_V$ is too big to
be consistent with the sum rules. On the other hand,
$G_S$ is needed to resolve a long standing
puzzle of QCD sum rules \cite{novikov}, that is,
the failure of the standard approach \cite{SVZ} in the pseudoscalar
channel.

It might worth emphasizing that the $G_{S,V}$ play different roles
in building up the hadron phenomenology. Roughly speaking, $G_S$
is needed to
generate spontaneous breaking of the chiral symmetry and to explain
existence of nearly massless pion in accordance with the
original idea \cite{nambu}. Introduction of $G_V$ allows
us to reproduce the vector meson dominance. Our conclusions
are that while the latter function of the extended NJL model can
hardly be
recoinciled with the QCD, the effective interaction (\ref{one})
with (\ref{numbers})
in the pseudoscalar channel
can be a missing block of the QCD-based phenomenology.

Let us recall the reader the general features of the sum rules
\cite{SVZ}.
The sum rules in the $\rho$ channel
are formulated in terms of $\Pi^{\rho} (Q^2)$,
\beq
\Pi^{\rho}(Q^2)(q_{\mu}q_{\nu}-g_{\mu\nu}q^2)~=~i\int d^4xe^{iqx}
\langle 0|
T\{j_{\mu}^{\rho}(x),j_{\nu}^{\rho}(0)\}
| 0\rangle,~~Q^2\equiv-q^2,
\label{corr}
\eeq
where $j_{\mu}^{\rho}$ is the quark current with quantum numbers of the
$\rho$ meson:
$$
j_{\mu}^{\rho}~=~{1\over 2}(\bar u\gamma_{\mu}u-\bar d\gamma_{\mu}d).
$$
The function $\Pi^{\rho}(Q^2)$ satisfies once subtracted dispersion
relations
\beq
\Pi^{\rho} (Q^2)~=~{{Q^2}\over{\pi}}\int{
{{R^{I=1}(s)ds}\over {s(s+Q^2)}} },
\eeq
where $R^{I=1}(s)$ is the ratio of the cross section of $e^+e^-$
annihilation
into hadrons with total isospin $I=1$ to that of annihilation into
$\mu^+\mu^-$ pair; in particular, $R^{I=1}(s)$ is contributed
by the $\rho$ meson.

The basic idea of the QCD sum rules \cite{SVZ} is
to calculate $\Pi^{\rho}(Q^2)$ at
large  $Q^2$ using the perturbative QCD and then extrapolate the result
to as
low $Q^2$ as possible. Moreover, it turns out that the advance towards
lowest
$Q^2$ is checked by corrections proportional to powers of $Q^{-2}$.
The coefficients in front of the powers of $Q^{-2}$ are related to
the quark condensate $\langle 0|\bar q q
|0\rangle$ and the gluon condensate $\langle 0|
G_{\mu\nu}^aG_{\mu\nu}^a|0\rangle$. Numerically,
the power corrections set
in around $Q^2\sim m^2_{\rho}$. More precisely, the correction are
relatively
small at such $Q^2$ but blow up fast at lower $Q^2$.
Moreover the sum rules are
most successful once applied to the Borel transform $\Pi^{\rho} (M^2)$
of $\Pi^{\rho} (Q^2)$ defined as
\beq
\Pi^{\rho} (M^2)~\equiv ~\hat{L}\Pi^{
\rho}(Q^2)~=~{\rm lim}_{n\rightarrow \infty}
{1 \over (n-1)!} (-1)^n(Q^2)^n({d\over {dQ^2}})^n
\Pi^{\rho} (Q^2) \label{borel},
\eeq
where the limit is understood in such a way that
\beq
n\rightarrow \infty,~~Q^2\rightarrow\infty,~~Q^2/n
\equiv M^2~{\rm is~fixed},
\eeq
and it is $M^2\sim m^2_{\rho}$ rather than $Q^2\sim m^2_{\rho}$
that can be reached starting from large $M^2$.

In a somewhat simplified form the sum rules read
\beq
1+\left({{0.2~{\rm GeV}^2}\over {M^2}}\right)^2-
\left({{0.3~{\rm GeV}^2}\over{M^2}}\right)^
3+O(M^{-8})
\label{sr}\eeq
$$~~~~~=
{{8\pi^2m^2_{\rho}}\over{g_{\rho}^2 M^2}}exp(-m^2_{\rho}/M^2)
+{2\over 3M^2}
\int_{s_0}^{\infty}exp(-s/M^2)R^{I=1}(s)ds,$$
where $M$ is a variable,
$g_{\rho}$ is the $\rho$ meson coupling related to the $e^+e^-$ width of
the $\rho$ meson, $g_{\rho}^2/4\pi \approx 3$, and
the integral on the right-hand side represents contribution of the
continuum.
For our purposes it is crucial only that even at $M^2
\approx m^2_{\rho}$ the
left-hand side of the sum rules (\ref{sr}) is calculable within the
short-distance approach to QCD, i.e., is dominated by the unit
corresponding to the bare loop graph. The sum rules agree with the
data, or
the right-hand side, to within about 10 per cent at $M^2
\sim 0.5~{\rm GeV}^2$.

Now we come to the crucial point of consistency of the effective
interaction
(\ref{one}) with the QCD sum rules. Adding
the contribution of interaction (1) to the bare quark loop, we get a
correction:
\beq
{{\delta\Pi^{\rho} (Q^2)}\over {\Pi^{\rho}_0(Q^2)}}~\approx~
-{{ G_VQ^2}\over {2\pi^2}}lnQ^2,
\eeq
or
\beq
{{\delta\Pi^{\rho}(M^2)}\over {\Pi^{\rho}_0(M^2)}}\approx
{{G_VM^2}\over{ \pi^2}}(1-\gamma+lnM^2),
\label{new}
\eeq
where $\Pi^{\rho}_0
$ is the contribution of the bare loop with
massless quarks and $\gamma~\approx~0.577$ is the Euler constant.
Note that the effect of the new interaction grows with $M^2$. On other
other
hand, at large $M^2$ it should disappear because of the onset of the
asymptotic freedom. This emphasizes once more that the effective
interaction
(\ref{one}) can be valid only at relatively low momenta. Once $M^2$
approaches $\Lambda_{UV}^2$, we have to allow for a form factor
due to the softening of the effective interaction. However,
at $M^2\sim 0.5~{\rm GeV}^2$ the correction (\ref{new}) is
estimated reliably.

To be consistent with the sum rules the correction (\ref{new}) is to
be small
at $M^2\sim 0.5~{\rm GeV}^2$. However, if we substitute the numbers,
then
we find in fact $\delta\Pi^{\rho}(0.5 ~{\rm GeV}^2)\sim \Pi^{\rho}_0
(0.5~{\rm GeV}^2)$.
Thus the effect of the new interaction cannot be even treated as a
correction and should be iterated.
Thus, we conclude that either the value of $G_V$ or of the cutoff
$\Lambda_{UV}$ given by $(\ref{numbers})$ is too high.
We will argue next that the consideration of the sum rules
in the pion channel favors the former possibility.

Now, we consider sum rules in the $\pi$ channel in exactly the same
way as
in the $\rho$ channel outlined above. The corresponding current is
defined as
\beq
j^{\pi}~=~{1\over 2}(\bar u i\gamma_5u-\bar d i\gamma_5d),
\eeq
and we introduce $\Pi^{\pi}$ in terms of the correlator similar
to (\ref{corr}) but without factorizing $(q^2g_{\mu\nu}-q_{\mu}
q_{\nu})$. The sum rules take the form \cite{novikov}:
\beq
\left({{\alpha_s(M^2)}\over{\alpha_s(\mu^2)}}\right)^{8/9}
\left(1+\left({{0.3~{\rm GeV}^2}\over {M^2}}\right)^2+2\left(
{{0.2~{\rm GeV}^2}\over {M^2}}\right)^3+O(M^{-8})\right)\label{pi}
\eeq
$$={{16\pi}\over{3M^4}}\int ds~exp(~-s/M^2)Im \Pi^{\pi}(s),$$
where
$Im\Pi^{\pi}(s)$ is the imaginary part of the correlator of two currents
$j^{\pi}$, the factor
 $(\alpha_s(M^2)$ $/\alpha_s(\mu^2))^{8/9}$
 is
due to a non-vanishing anomalous dimension of $j^{\pi}$
and $\alpha_s$ is the running QCD coupling. As far as
$Im\Pi^{\pi}(s)$ is concerned, the only well-known contribution to
it comes from
the pion:
\beq
Im\Pi^{\pi}_{\rm pole}~=~\pi f_{\pi}^2m_{\pi}^4(m_u+m_d)^{-2}
\delta(s-m^2_{\pi}),
\label{pole}\eeq
where $f_{\pi}\approx 93~ {\rm MeV}$ and $m_{u,d}$ are the current
quark masses.

Now, it has been demonstrated that the sum rules (\ref{pi}) do not hold
experimentally as they are stated \cite{novikov}. The point is that the
pole contribution (\ref{pole}) alone, with negligence of the rest
of $Im \Pi^{\pi}(s)$ which is positive definite, is too large to be
consistent with (\ref{pi}). More specifically, one can prove
existence of a new contribution, unaccounted in the sum rules,
for $M^2$ ranged between
$0.5~{\rm GeV}^2<M^2<2~{\rm GeV}^2$.
At $M^2\sim 0.5~{\rm GeV}^2$  the new contribution is no less
than that of the pion, while at $M^2\sim~ 2~{\rm GeV}^2$ it is still
larger than
10 per cent of the pion contribution.
The sensitivity of the sum rules at lower $M^2$ is limited
by the power corrections and at larger $M^2$ the bare loop may
dominate.

What we propose here is to
ascribe this new contribution to the effective interaction (\ref{one})
in the $\pi$ channel.
Consider first the $G_S$ term as a perturbation. Then
\beq
{{\delta \Pi^{\pi} (Q^2)}\over {\Pi^{\pi}_0 (Q^2)}}~\approx~
{{3G_SQ^2}\over {4\pi^2}}lnQ^2,
\eeq
or
\beq
{{\delta\Pi^{\pi}(M^2)}\over{\Pi^{\pi}_0(M^2)}}~\approx~
-{{3G_SM^2}\over{\pi^2}}\left({3\over 2} -\gamma+lnM^2\right).
\label{large}
\eeq
This equation is supposed to be valid at $M^2$ much smaller than
$\Lambda_{UV}^2$. Applying it at $M^2=0.5~{\rm GeV}^2$, we find:
\beq
{{\delta\Pi^{\pi}(M^2=0.5{\rm GeV}^2)}
\over{\Pi^{\pi}_0(M^2=0.5{\rm GeV}^2)}}~\sim ~2.5.
\eeq
Literally, the analysis of the sum rules suggests a new contribution
of order
one, in the same units. The factor we get now is in rough agreement
with this estimate.

To get a better estimate we should have iterated the effect of the
new interaction, since it turns out to be large. The change brought by
the iterations of the effective intertaction within the NJL model
is remarkably simple and
well known. Namely, within the the NJL model the summation
of the loops generated by the $G_S$ term produces a pion.
In this way one reproduces
the contribution of the pion into the right-hand side of eq.(\ref{pi})
and explains the failure of the sum rules which do not account for the
interaction (\ref{one}).

Thus the new contribution to the sum rules gets a natural explanation.
Namely, the four-fermion interaction gives rise to the pion as
proposed in the original papers \cite{nambu}.
At $M^2\sim\Lambda_{UV}^2\sim 1.5~{\rm GeV}^2$ the effective
interaction is dissolved and the sum rules get dominated by the bare
quark loop graph. The estimate of the mass scale of the onset of
asymptotic freedom in the pion channel as 2~${\rm GeV}^2$ (see above)
turns out to
be in reasonable agreement with the estimate of $\Lambda_{UV}^2$ within
the NJL model.

This picture by itself does not explain the difference between the
vector and the pseudoscalar channels. There should be a new kind of
correction
to the sum rules \cite{novikov} within the fundamental QCD.
 If the new correction is a $1/M^2$
term \cite{VZrecent} associated with the ultraviolet renormalon, then
the difference between the channels is not so dramatic, say, a factor
of 4 would suffice. This interpretation is also supported by the
recent observation that ultraviolet renormalon is associated with
four-quark
interactions \cite{vainshtein}. However, there is no reliable way to
find the relative weight of $G_S$ and $G_V$.

The chain of the arguments get closed through the prediction
obtained above,
\beq
G_V~\ll~G_S, \label{lll}
\eeq
which can be tested within the NJL model. Inspection of the most recent
fits \cite{bijnens} reveals independent evidence in favor
of (\ref{lll}).
In particular, any $G_V\neq 0$ drives the predicted
value of the constant $g_A$ governing the beta-decay of neutron off
its experimental value. Less dramatically, taking $G_V=0$ improves
agreement with the data in some other cases as well. Furthermore,
as is noted in \cite{bijnens} the NJL model with $G_V=0$ is
equivalent to the effective QCD Lagrangian of Ref.\cite{espriu}
which describes successfully nonleptonic weak decays and
electromagnetic properties of the pion \cite{pich}.

Thus, it seems fair to say that the NJl model with
$G_V=0,~G_S\neq 0$ results in a sound phenomenology, although it puts
$\pi$ and $\rho$ mesons on different footing.

The authors are thankful to G. Grunberg and M.K. Volkov for useful
discussions.
One of the authors (V.I.Z.) acknowledges warm hospitality extended to
him
during his stay with the theory group at Nagoya University when this
work
was started. Part of this work was done while K.Y. was at Institute
for Theoretical Physics, University of California at Santa Barbara
(Workshop "Weak Interactions").
The work was supported in part by the U.S. Department of Energy, and
by a Grant-in-Aid for Scientific Research
from the Japanese Ministry of Education, Science and Culture
(No. 05640339) and the Ishida Foundation.

\enddocument

\end{document}